# The Anatomy of the Grid
## Enabling Scalable Virtual Organizations *

Ian Foster •¶    Carl Kesselman §    Steven Tuecke •

{foster, tuecke}@mcs.anl.gov, carl@isi.edu

**Abstract**

"Grid" computing has emerged as an important new field, distinguished from conventional distributed computing by its focus on large-scale resource sharing, innovative applications, and, in some cases, high-performance orientation. In this article, we define this new field. First, we review the "Grid problem," which we define as flexible, secure, coordinated resource sharing among dynamic collections of individuals, institutions, and resources—what we refer to as *virtual organizations*. In such settings, we encounter unique authentication, authorization, resource access, resource discovery, and other challenges. It is this class of problem that is addressed by Grid technologies. Next, we present an extensible and open *Grid architecture*, in which protocols, services, application programming interfaces, and software development kits are categorized according to their roles in enabling resource sharing. We describe requirements that we believe any such mechanisms must satisfy and we discuss the central role played by the *intergrid protocols* that enable interoperability among different Grid systems. Finally, we discuss how Grid technologies relate to other contemporary technologies, including enterprise integration, application service provider, storage service provider, and peer-to-peer computing. We maintain that Grid concepts and technologies complement and have much to contribute to these other approaches.

## 1   Introduction

The term "the Grid" was coined in the mid1990s to denote a proposed distributed computing infrastructure for advanced science and engineering [31]. Considerable progress has since been made on the construction of such an infrastructure (e.g., [10, 15, 43, 55]), but the term "Grid" has also been conflated, at least in popular perception, to embrace everything from advanced networking to artificial intelligence. One might wonder whether the term has any real substance and meaning. Is there really a distinct "Grid problem" and hence a need for new "Grid technologies"? If so, what is the nature of these technologies, and what is their domain of applicability? While numerous groups have interest in Grid concepts and share, to a significant extent, a common vision of Grid architecture, we do not see consensus on the answers to these questions.

Our purpose in this article is to argue that the Grid concept is indeed motivated by a real and specific problem and that there is an emerging, well-defined Grid technology base that solves this problem. In the process, we develop a detailed architecture and roadmap for current and future Grid technologies. Furthermore, we assert that while Grid technologies are currently distinct from other major technology trends, such as Internet, enterprise, distributed, and peer-to-peer computing, these other trends can benefit significantly from growing into the problem space addressed by Grid technologies.

• Mathematics and Computer Science Division, Argonne National Laboratory, Argonne, IL 60439.
¶ Department of Computer Science, The University of Chicago, Chicago, IL 60657.
§ Information Sciences Institute, The University of Southern California, Marina del Rey, CA 90292.

The real and specific problem that underlies the Grid concept is *coordinated resource sharing and problem solving in dynamic, multi-institutional virtual organizations*. The sharing that we are concerned with is not primarily file exchange but rather direct access to computers, software, data, and other resources, as is required by a range of collaborative problem-solving and resource-brokering strategies emerging in industry, science, and engineering. This sharing is, necessarily, highly controlled, with resource providers and consumers defining clearly and carefully just what is shared, who is allowed to share, and the conditions under which sharing occurs. A set of individuals and/or institutions defined by such sharing rules form what we call a *virtual organization* (VO).

The following are examples of VOs: the application service providers, storage service providers, cycle providers, and consultants engaged by a car manufacturer to perform scenario evaluation during planning for a new factory; members of an industrial consortium bidding on a new aircraft; a crisis management team and the databases and simulation systems that they use to plan a response to an emergency situation; and members of a large, international, multiyear high-energy physics collaboration. Each of these examples represents an approach to computing and problem solving based on collaboration in computation- and data-rich environments.

As these examples show, VOs vary tremendously in their purpose, scope, size, duration, structure, community, and sociology. Nevertheless, careful study of underlying technology requirements leads us to identify a broad set of common concerns and requirements. In particular, we see a need for highly flexible sharing relationships, ranging from client-server to peer-to-peer and brokered; for complex and high levels of control over how shared resources are used, including fine-grained access control, delegation, and application of local and global policies; for sharing of varied resources, ranging from programs, files, and data to computers, sensors, and networks; and for diverse usage modes, ranging from single user to multi-user and from performance sensitive to cost-sensitive and hence embracing issues of quality of service, scheduling, co-allocation, and accounting.

Current distributed computing technologies do not address the concerns and requirements just listed. For example, current Internet technologies address communication and information exchange among computers but not the coordinated use of resources at multiple sites for computation. Business-to-business exchanges [53] focus on information sharing (often via centralized servers). So do virtual enterprise technologies, although here sharing may eventually extend to applications and physical devices (e.g., [8]). Enterprise distributed computing technologies such as CORBA and Enterprise Java focus on enabling resource sharing within a single organization. Storage service providers (SSPs) and application service providers (ASPs) allow organizations to outsource storage and computing requirements to other parties, but only in constrained ways: for example, SSP resources are typically linked with a customer via a virtual private network (VPN). Emerging "Internet computing" companies seek to harness idle computers on an international scale [28] but, to date, support only highly centralized access to those resources. In summary, current technology either does not accommodate the range of resource types or does not provide the flexibility and control on sharing relationships needed to establish VOs.

It is here that Grid technologies enter the picture. Over the past five years, research and development efforts within the Grid community have produced protocols, services, and tools that address precisely the challenges that arise when we seek to build scalable VOs. These technologies include security solutions that support management of credentials and policies when computations span multiple institutions; resource management protocols and services that support secure remote access to computing and data resources and the co-allocation of multiple resources; information query protocols and services that provide configuration and status information about resources, organizations, and services; and data management services that locate and transport datasets between storage systems and applications.

Because of their focus on dynamic, cross-organizational sharing, Grid technologies complement rather than compete with existing distributed computing technologies.  For example, enterprise distributed computing systems can use Grid technologies to achieve resource sharing across institutional boundaries; in the ASP/SSP space, Grid technologies can be used to establish dynamic markets for computing and storage resources, hence overcoming the limitations of current static configurations.  We discuss the relationship between Grids and these technologies in more detail below.

In the rest of this article, we expand upon each of these points in turn.  Our objectives are to (1) clarify the nature of VOs and Grid computing for those unfamiliar with the area; (2) contribute to the emergence of Grid computing as a discipline by establishing a standard vocabulary and defining an overall architectural framework; and (3) define clearly how Grid technologies relate to other technologies, explaining both why various emerging technologies are not yet the Grid and how these technologies can benefit from Grid technologies.

It is our belief that VOs have the potential to change dramatically the way we use computers to solve problems, much as the web has changed how we exchange information.  As the examples presented here illustrate, the need to engage in collaborative processes is fundamental to many diverse disciplines and activities: it is not limited to science, engineering and business activities.  It is because of this broad applicability of VO concepts that Grid technology is important.

## 2   The Emergence of Virtual Organizations

Consider the following four scenarios:

1. A company needing to reach a decision on the placement of a new factory invokes a sophisticated financial forecasting model from an ASP, providing it with access to appropriate proprietary historical data from a corporate database on storage systems operated by an SSP.  During the decision-making meeting, what-if scenarios are run collaboratively and interactively, even though the division heads participating in the decision are located in different cities.  The ASP itself contracts with a cycle provider for additional "oomph" during particularly demanding scenarios, requiring of course that cycles meet desired security and performance requirements.

2. An industrial consortium formed to develop a feasibility study for a next-generation supersonic aircraft undertakes a highly accurate multidisciplinary simulation of the entire aircraft.  This simulation integrates proprietary software components developed by different participants, with each component operating on that participant's computers and having access to appropriate design databases and other data made available to the consortium by its members.

3. A crisis management team responds to a chemical spill by using local weather and soil models to estimate the spread of the spill, determining the impact based on population location as well as geographic features such as rivers and water supplies, creating a short-term mitigation plan (perhaps based on chemical reaction models), and tasking emergency response personnel by planning and coordinating evacuation, notifying hospitals, and so forth.

4. Thousands of physicists at hundreds of laboratories and universities worldwide come together to design, create, operate, and analyze the products of a major detector at CERN, the European high energy physics laboratory.  During the analysis phase, they pool their computing, storage, and networking resources to create a "Data Grid" capable of analyzing petabytes of data [21, 41, 50].

These four examples differ in many respects: the number and type of participants, the types of activities, the duration and scale of the interaction, and the resources being shared. But they also have much in common, as discussed in the following (see also Figure 1).

In each case, a number of mutually distrustful participants with varying degrees of prior relationship (perhaps none at all) want to share resources in order to perform some task. Furthermore, sharing is about more than simply document exchange (as in "virtual enterprises" [17]): it can involve direct access to remote software, computers, data, and other resources. For example, members of a consortium may provide access to specialized software and data and/or pool their computational resources.

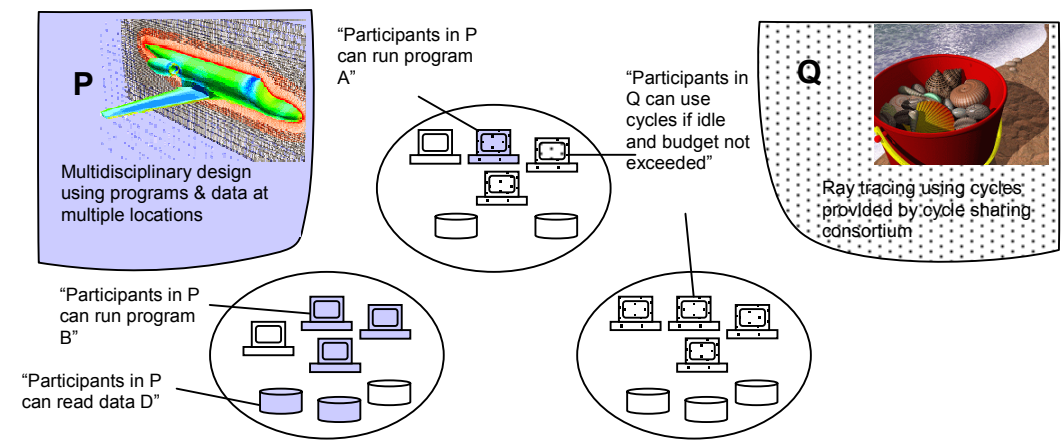

**Figure 1**: An actual organization can participate in one or more VOs by sharing some or all of the resources that the actual organization controls. We show three actual organizations (the circles), and two VOs: P, which links participants in an aerospace design consortium, and Q, which links colleagues who have agreed to share space cycles. One actual organization participates in P, the second participates in Q, and the third is a member of both P and Q. The policies governing access to resources (summarized in "quotes") vary according to the actual organizations, resources, and VOs involved.

Resource sharing is conditional: each resource owner makes resources available subject to—often stringent—constraints on when, where, and what can be done. For example, a participant in VO P of Figure 1 might allow VO partners to invoke their simulation service only for "simple" problems. Resource consumers may also place constraints on properties of the resources they are prepared to work with. For example, a participant in VO Q might accept only pooled computational resources certified as "secure." The implementation of such constraints requires mechanisms for expressing policies, for establishing the identity of a consumer or resource (authentication), and for determining whether an operation is consistent with applicable sharing relationships (authorization).

Sharing relationships can vary dynamically over time, in terms of the resources involved, the nature of the access permitted, and the participants to whom access is permitted. And these relationships do not necessarily involve an explicitly named set of individuals, but rather may be defined implicitly by the policies that govern access to resources. For example, an organization might enable access by anyone who can demonstrate that they are a "customer" or a "student."

The dynamic nature of sharing relationships means that we require mechanisms for discovering and characterizing the nature of the relationships that exist at a particular point in time. For example, a new participant joining VO Q must be able to determine what resources it is able to access, the "quality" of these resources, and the policies that govern access.

Sharing relationships are often not simply client-server, but peer to peer: providers can be consumers, and sharing relationships can exist among any subset of participants. Sharing relationships may be combined to coordinate use across many resources, each owned by different organizations. For example, in VO Q, a computation started on one pooled computational resource may subsequently access data or initiate subcomputations elsewhere. The ability to delegate authority in controlled ways becomes important in such situations, as do mechanisms for coordinating operations across multiple resources (e.g., coscheduling).

The same resource may be used in different ways, depending on the restrictions placed on the sharing and the goal of the sharing. For example, a computer may be used only to run a specific piece of software in one sharing arrangement, while it may provide generic compute cycles in another. Because of the lack of a priori knowledge about how a resource may be used, performance metrics, expectations, and limitations (i.e., quality of service) may be part of the conditions placed on resource sharing or usage.

These characteristics and requirements define what we term a *virtual organization*, a concept that we believe is becoming fundamental to much of modern computing. VOs enable disparate groups of organizations and/or individuals to share resources in a controlled fashion, so that members may collaborate to achieve a shared goal.

## 3  The Nature of Grid Architecture

The establishment, management, and exploitation of cross-organizational VO sharing relationships require new technology. We structure our discussion of this technology in terms of a *Grid architecture* that identifies fundamental system components, specifies the purpose and function of these components, and indicates how these components interact with one another.

In defining a Grid architecture, we start from the perspective that effective VO operation requires that we be able to establish sharing relationships among *any* potential participants. Interoperability is thus the central issue to be addressed. In a networked environment, interoperability means common protocols. Hence, our Grid architecture is first and foremost a *protocol* architecture, with protocols defining the basic mechanisms by which VO users and resources negotiate, establish, manage, and exploit sharing relationships. A standards-based open architecture facilitates extensibility, interoperability, portability, and code sharing; standard protocols make it easy to define standard services that provide enhanced capabilities. We can also construct Application Programming Interfaces and Software Development Kits (see Appendix for definitions) to provide the programming abstractions required to create a usable Grid. Together, this technology and architecture constitute what is often termed middleware ("the services needed to support a common set of applications in a distributed network environment" [3]), although we avoid that term here due to its vagueness. We discuss each of these points in the following.

Why is interoperability such a fundamental concern? At issue is our need to ensure that sharing relationships can be initiated among arbitrary parties, accommodating new participants dynamically, across different platforms, languages, and programming environments. In this context, mechanisms serve little purpose if they are not defined and implemented so as to be interoperable across organizational boundaries, operational policies, and resource types. Without interoperability, VO applications and participants are forced to enter into bilateral sharing arrangements, as there is no assurance that the mechanisms used between any two parties will extend to any other parties. Without such assurance, dynamic VO formation is all but impossible, and the types of VOs that can be formed are severely limited. Just as the Web revolutionized information sharing by providing a universal protocol and syntax (HTTP and HTML) for information exchange, so we require standard protocols and syntaxes for general resource sharing.

Why are protocols critical to interoperability? A protocol definition specifies how distributed system elements interact with one another in order to achieve a specified behavior, and the structure of the information exchanged during this interaction. This focus on externals (interactions) rather than internals (software, resource characteristics) has important pragmatic benefits. VOs tend to be fluid; hence, the mechanisms used to discover resources, establish identity, determine authorization, and initiate sharing must be flexible and lightweight, so that resource-sharing arrangements can be established and changed quickly. Because VOs complement rather than replace existing institutions, sharing mechanisms cannot require substantial changes to local policies and must allow individual institutions to maintain ultimate control over their own resources. Since protocols govern the interaction between components, and not the implementation of the components, local control is preserved.

Why are services important? A service (see Appendix) is defined solely by the protocol that it speaks and the behaviors that it implements. The definition of standard services—for access to computation, access to data, resource discovery, coscheduling, data replication, and so forth—allows us to enhance the services offered to VO participants and also to abstract away resource-specific details that would otherwise hinder the development of VO applications.

Why do we also consider Application Programming Interfaces (APIs) and Software Development Kits (SDKs)? There is, of course, more to VOs than interoperability, protocols, and services. Developers must be able to develop sophisticated applications in complex and dynamic execution environments. Users must be able to operate these applications. Application robustness, correctness, development costs, and maintenance costs are all important concerns. Standard abstractions, APIs, and SDKs can accelerate code development, enable code sharing, and enhance application portability. APIs and SDKs are an adjunct to, not an alternative to, protocols. Without standard protocols, interoperability can be achieved at the API level only by using a single implementation everywhere—infeasible in many interesting VOs—or by having every implementation know the details of every other implementation. (The Jini approach [6] of downloading protocol code to a remote site does not circumvent this requirement.)

In summary, our approach to Grid architecture emphasizes the identification and definition of protocols, first; services, second; and APIs and SDKs, third.

## 4 Grid Architecture Description

Our goal in describing our Grid architecture is not to provide a complete enumeration of all required protocols (and services, APIs, and SDKs) but rather to identify requirements for general classes of component. The result is an extensible, open architectural structure within which can be placed solutions to key VO requirements. Our architecture and the subsequent discussion organize components into layers, as shown in Figure 2. Components within each layer share common characteristics but can build on capabilities and behaviors provided by any lower layer.

In specifying the various layers of the Grid architecture, we follow the principles of the "hourglass model" [1]. The neck of the hourglass defines a fundamental set of core abstractions and protocols, onto which many different high-level behaviors can be mapped (the top of the hourglass), and which themselves can be mapped onto many different underlying technologies (the base of the hourglass). By definition, the number of protocols defined at the neck must be small. In our architecture, the neck of the hourglass consists of *Resource* and *Connectivity* protocols, which facilitate the sharing of individual resources. Protocols at these layers are designed so that they can be implemented on top of a diverse range of resource types, defined at the *Fabric* layer, and can in turn be used to construct a wide range of global services and application-specific behaviors at the *Collective* layer.

Our architectural description is high level and places few constraints on design and implementation. To make this abstract discussion more concrete, we also list, *for illustrative purposes*, the protocols defined within the Globus Toolkit [30], and used within such Grid projects as the NSF's National Technology Grid [55], NASA's Information Power Grid [43], DOE's DISCOM [10], and the European Data Grid. More details are provided elsewhere [33].

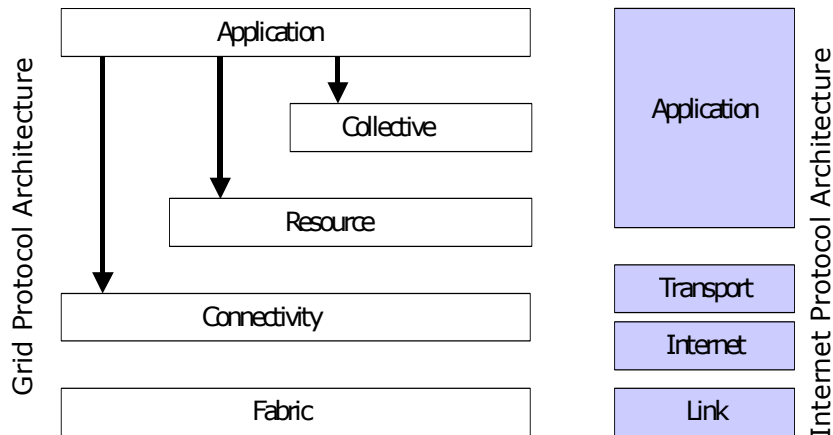

**Figure 2**: The layered Grid architecture and its relationship to the Internet protocol architecture. Because the Internet protocol architecture extends from network to application, there is a mapping from Grid layers into Internet layers.

## 4.1 Fabric: Interfaces to Local Control

The Grid *Fabric* layer provides the resources to which shared access is mediated by Grid protocols: for example, computational resources, storage systems, catalogs, network resources, and sensors. A "resource" may be a logical entity, such as a distributed file system, computer cluster, or distributed computer pool; in such cases, a resource implementation may involve internal protocols (e.g., the NFS storage access protocol or a cluster resource management system's process management protocol), but these are not the concern of Grid architecture.

Fabric components implement the local, resource-specific operations that occur on specific resources (whether physical or logical) as a result of sharing operations at higher levels. There is thus a tight and subtle interdependence between the functions implemented at the Fabric level, on the one hand, and the sharing operations supported, on the other. Richer Fabric functionality enables more sophisticated sharing operations; at the same time, if we place few demands on Fabric elements, then deployment of Grid infrastructure is simplified. For example, if resources support advance reservations, then it is straightforward to implement higher-level services that coschedule multiple resources. However, as in practice few resources support advance reservation "out of the box," a requirement for advance reservation increases the cost of incorporating new resources into a Grid.

Experience suggests that at a minimum, resources should implement *enquiry* mechanisms that permit discovery of their structure and state, on the one hand, and *resource management* mechanisms that provide some control of delivered quality of service, on the other. The following brief and partial list provides a resource-specific characterization of capabilities.

- *Computational resources*: Mechanisms are required for starting programs and for monitoring and controlling the execution of the resulting processes. Management mechanisms that allow control over the resources allocated to processes are useful, as are advance reservation mechanisms. Enquiry functions are needed for determining hardware and software characteristics as well as relevant load information such as current load and queue state in the case of scheduler-managed resources.

- *Storage resources*: Mechanisms are required for putting and getting files. Third-party and high-performance (e.g., striped) transfers are useful [57]. So are mechanisms for reading and writing subsets of a file and/or executing remote data selection or reduction functions [14]. Management mechanisms that allow control over the resources allocated to data transfers (space, disk bandwidth, network bandwidth, CPU) are useful, as are advance reservation mechanisms. Enquiry functions are needed for determining hardware and software characteristics as well as relevant load information such as available space and bandwidth utilization.

- *Network resources*: Management mechanisms that provide control over the resources allocated to network transfers (e.g., prioritization, reservation) can be useful. Enquiry functions should be provided to determine network characteristics and load.

- *Code repositories*: This specialized form of storage resource requires mechanisms for managing versioned source and object code: for example, a control system such as CVS.

- *Catalogs*: This specialized form of storage resource requires mechanisms for implementing catalot query and update operations: for example, a relational database [9].

*Globus Toolkit*: The Globus Toolkit has been designed to use (primarily) existing fabric components, including vendor-supplied protocols and interfaces. If the necessary Fabric-level behavior is not provided by a vendor, however, the Globus Toolkit includes the missing functionality. For example, enquiry software is provided for discovering structure and state information for various common resource types, such as computers (e.g., OS version, hardware configuration, load [27], scheduler queue status), storage systems (e.g., available space), and networks (e.g., current and predicted future load [49, 59]), and for packaging this information in a form that facilitates the implementation of higher-level protocols, specifically at the Resource layer. Resource management, on the other hand, is generally assumed to be the domain of local resource managers. One exception is the General-purpose Architecture for Reservation and Allocation (GARA) [34], which provides a "slot manager" that can be used to implement advance reservation for resources that do not support this capability. Others have developed enhancements to the Portable Batch System (PBS) [52] and Condor [46, 47] that support advance reservation capabilities.

## *4.2 Connectivity: Communicating Easily and Securely*

The *Connectivity* layer defines core communication and authentication protocols required for Grid-specific network transactions. Communication protocols enable the exchange of data between Fabric layer resources. Authentication protocols build on communication services to provide cryptographically secure mechanisms for verifying the identity of users and resources.

Communication requirements include transport, routing, and naming. While alternatives certainly exist, in almost all practical situations these protocols will be drawn from the TCP/IP protocol stack: specifically, the Internet (IP and ICMP), transport (TCP, UDP), and application (DNS, OSPF, RSVP, etc.) layers of the Internet layered protocol architecture [7].

With respect to security aspects of the Connectivity layer, we observe that the complexity of the security problem makes it important that any solutions be based on existing standards whenever possible. As with communication, many of the security standards developed within the context of the Internet protocol suite are applicable.

Authentication solutions for VO environments should have the following characteristics [16]:

- *Single sign on*. Users must be able to "log on" (authenticate) just once and then have access to multiple Grid resources defined in the Fabric layer, without further user intervention.

- *Delegation* [32, 37, 42]. A user must be able to endow a program with the ability to run on that user's behalf, so that the program is able to access the resources on which the user is authorized. The program should (optionally) also be able to conditionally delegate a subset of its rights to another program (sometimes referred to as restricted delegation).

- *Integration with various local security solutions*: Each site or resource provider may employ any of a variety of local security solutions, including Kerberos and Unix security. Grid security solutions must be able to interoperate with these various local solutions. They cannot, realistically, require wholesale replacement of local security solutions but rather must allow mapping into the local environment.

- *User-based trust relationships*: In order for a user to use resources from multiple providers together, the security system must not require each of the resource providers to cooperate or interact with each other in configuring the security environment. For example, if a user has the right to use sites A and B, the user should be able to use sites A and B together without requiring that A's and B's security administrators interact.

Grid security solutions should also provide flexible support for communication protection (e.g., control over the degree of protection, independent data unit protection for unreliable protocols, support for reliable transport protocols other than TCP) and enable stakeholder control over authorization decisions, including the ability to restrict the delegation of rights in various ways.

*Globus Toolkit*: The Internet protocols are used for communication. The public-key based Grid Security Infrastructure (GSI) protocols [16, 32] are used for authentication, communication protection, and authorization. GSI builds on and extends the Transport Layer Security (TLS) protocols [26] to address most of the issues listed above: in particular, single sign-on, delegation, integration with various local security solutions (including Kerberos [54]), and user-based trust relationships. X.509-format certificates are used. Stakeholder control of authorization is supported via an authorization toolkit that allows resource owners to integrate local policies via a Generic Authorization and Access (GAA) control interface. Rich support for restricted delegation is not provided in the production toolkit but has been demonstrated in prototypes.

## 4.3  Resource: Sharing Single Resources

The Resource layer builds on Connectivity layer communication and authentication protocols to define protocols (and APIs and SDKs) for the secure initiation, monitoring, and control of sharing operations on individual resources. Resource layer implementations of these protocols call Fabric layer functions to access and control local resources. Resource layer protocols are concerned entirely with individual resources and hence ignore issues of global state and atomic actions across distributed collections; such issues are the concern of the Collective layer discussed next.

Two primary classes of Resource layer protocols can be distinguished:

- *Information protocols* are used to obtain information about the structure and state of a resource, for example, its configuration, current load, and usage policy.

- *Management protocols* are used to negotiate access to a shared resource, specifying, for example, resource requirements (including advanced reservation and quality of service) and the operation(s) to be performed, such as process creation, or data access. Since management protocols are responsible for instantiating sharing relationships, they must serve as a "policy application point," ensuring that the requested protocol operations are consistent with the policy under which the resource is to be shared. Issues that must be considered include accounting and payment. A protocol may also support monitoring the status of an operation and controlling (for example, terminating) the operation.

While many such protocols can be imagined, the Resource (and Connectivity) protocol layers form the neck of our hourglass model, and as such, we require a small and standard set. These protocols must be chosen so as to capture the fundamental mechanisms of sharing across many different resource types (for example, different local resource management systems), while not overly constraining the types or performance of higher-level protocols that may be developed.

The list of desirable Fabric functionality provided in Section 4.1 summarizes the major features required in Resource layer protocols. To this list we can add a need for "exactly once" semantics for many operations, with reliable error reporting indicating when operations fail.

*Globus Toolkit*: A small and mostly standards-based set of protocols is adopted. In particular:

- The Lightweight Directory Access Protocol (LDAP) is used to define a standard resource information protocol and associated information model. An associated soft-state resource registration protocol, the Grid Resource Registration Protocol (GRRP), is used to register resources with Grid Index Information Servers, discussed in the next section.

- The HTTP-based Grid Resource Access and Management (GRAM) protocol is used for allocation of computational resources and for monitoring and control of computation on those resources.

- An extended version of the File Transfer Protocol, GridFTP, is used for data access; extensions include use of Connectivity layer security protocols, partial file access, and management of parallelism for high-speed transfers. FTP is adopted as a base data transfer protocol because of its support for third-party transfers and because its separate control and data channels facilitate the implementation of sophisticated servers.

- LDAP is also used as a catalog access protocol.

The Globus Toolkit defines client-side C and Java APIs and SDKs for each of these protocols. Server-side SDKs and servers are also provided for each protocol, to facilitate the integration of various resources (computational, storage, network) into the Grid. For example, the Grid Resource Information Service (GRIS) implements server-side LDAP functionality, with callouts allowing for publication of arbitrary resource information. An important server-side element of the overall Toolkit is the gatekeeper, which provides what is in essence a GSI-authenticated "inetd" that speaks the GRAM protocol and can be used to dispatch various local operations. The Generic Security Services (GSS) API [45] is used for all authentication operations within these SDKs and servers, enabling substitution of alternative security services at the Connectivity layer.

## *4.4 Collective: Coordinating Multiple Resources*

While the Resource layer is focused on interactions with a single resource, the next layer in the architecture contains protocols and services (and APIs and SDKs) that are not associated with any one specific resource but rather are global in nature and capture interactions across collections of resources. For this reason, we refer to the next layer of the architecture as the *Collective* layer. Because Collective components build on the narrow Resource and Connectivity layer "neck" in the protocol hourglass, they can implement a wide variety of sharing behaviors without placing new requirements on the resources being shared. For example:

- *Directory services* allow VO participants to discover the existence and/or properties of VO resources. A directory service may allow its users to query for resources by name and/or by attributes such as type, availability, or load.

- *Co-allocation, scheduling, and brokering services* allow VO participants to request the allocation of one or more resources for a specific purpose and the scheduling of tasks on the appropriate resources. Examples include AppLeS [12, 13], Condor-G, Nimrod-G [2], and the DRM broker [10].

- *Monitoring and diagnostics services* support the monitoring of VO resources for failure, adversarial attack ("intrusion detection"), overload, and so forth.

- *Data replication services* support the management of VO storage (and perhaps also network and computing) resources to maximize data access performance with respect to metrics such as response time, reliability, and cost [4, 41].

- *Grid-enabled programming systems* enable familiar programming models to be used in Grid environments, using various Grid services to address resource discovery, security, resource allocation, and other concerns. Examples include Grid-enabled implementations of the Message Passing Interface [29, 35] and manager-worker frameworks [20, 38].

- *Software discovery services* discover and select the best software implementation and execution platform based on the parameters of the problem being solved [19]. Examples include NetSolve [18] and Ninf [51].

- *Community authorization servers* enforce community policies governing resource access, generating capabilities that community members can use to access community resources. These servers provide a global policy enforcement service by building on resource information, and resource management protocols (in the Resource layer) and security protocols in the Connectivity layer. Akenti [56] addresses some of these issues.

- *Collaboratory services* support the coordinated exchange of information within potentially large user communities, whether synchronously or asynchronously. Examples are CAVERNsoft [25, 44], Access Grid [22], and commodity groupware systems.

These examples illustrate the wide variety of Collective layer protocols and services that are encountered in practice. Notice that while Resource layer protocols must be general in nature and are widely deployed, Collective layer protocols span the spectrum from general purpose to highly application or domain specific, with the latter existing perhaps only within specific VOs.

Collective functions can be implemented as persistent services, with associated protocols, or as SDKs (with associated APIs) designed to be linked with applications. In both cases, their implementation can build on Resource layer (or other Collective layer) protocols and APIs. For example, Figure 3 shows a Collective co-allocation API and SDK (the middle tier) that uses a Resource layer management protocol to manipulate underlying resources. Above this, we define a co-reservation service protocol and implement a co-reservation service that speaks this protocol, calling the co-allocation API to implement co-allocation operations and perhaps providing additional functionality, such as authorization, fault tolerance, and logging. An application might then use the co-allocation service protocol to request end-to-end network reservations.

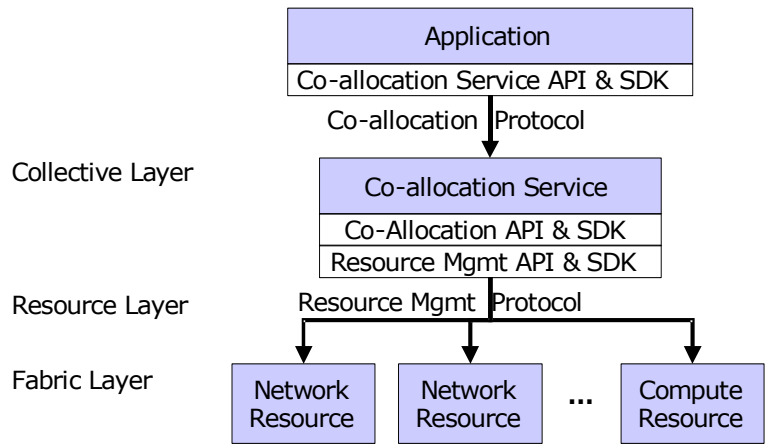

**Figure 3:** Collective and Resource layer protocols, services, APIs, and SDKS can be combined in a variety of ways to deliver functionality to applications.

Collective components may be tailored to the requirements of a specific user community, VO, or application domain, for example, an SDK that implements an application-specific coherency protocol, or a co-reservation service for a specific set of network resources. Other Collective components can be more general-purpose, for example, a replication service that manages an international collection of storage systems for multiple communities, or a directory service designed to enable the discovery of VOs. In general, the larger the target user community, the more important it is that a Collective component's protocol(s) and API(s) be standards based.

*Globus Toolkit*: In addition to the example services listed earlier in this section, many of which build on Globus Connectivity and Resource protocols, we mention the Meta Directory Service, which introduces Grid Information Index Servers (GIISs) to support arbitrary views on resource subsets, with the LDAP information protocol used to access resource-specific GRISs to obtain resource state and GRRP used for resource registration. Also replica catalog and replica management services used to support the management of dataset replicas in a Grid environment [4]. An online certificate repository service ("MyProxy") provides secure storage for proxy credentials. The DUROC co-allocation library provides an SDK and API for resource co-allocation [24].

## *4.5 Applications*

The final layer in our Grid architecture comprises the user applications that operate within a VO environment. Figure 4 illustrates an application programmer's view of Grid architecture. Applications are constructed in terms of, and by calling upon, services defined at any layer. At each layer, we have well-defined protocols that provide access to some useful service: resource management, data access, resource discovery, and so forth. At each layer, APIs may also be defined whose implementation (ideally provided by third-party SDKs) exchange protocol messages with the appropriate service(s) to perform desired actions.

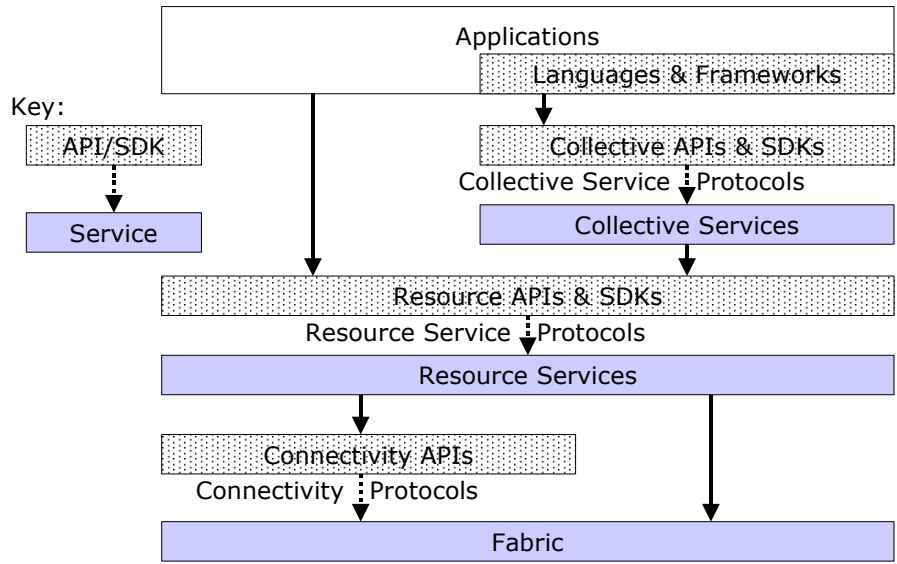

**Figure 4.** Software development kits (SDKs) implement specific APIs. These APIs in turn use Grid protocols to interact with network services that provide capabilities to the end user. Higher level SDKs can provide functionality that is not directly mapped to a specific protocol, but may combine protocol operations with calls to additional APIs as well as implement local functionality.

Notice the additional "Languages and Frameworks" component introduced in Figure 4. While the preceding discussion has focused on protocols as a means of achieving interoperability and APIs as a way of promoting code sharing and portability, effective application development can often benefit from the use of higher-level languages and frameworks (e.g., the Common Component Architecture [5], SciRun [19], CORBA [36, 48], Legion [39], Cactus [11]). These higher-level systems can build on protocols, services, and APIs provided within the Grid architecture.

## 5  Grid Architecture in Practice

We use two examples to illustrate how Grid architecture functions in practice. Table 1 shows the services that might be used to implement the multidisciplinary simulation and ray tracing applications introduced in Figure 1. The basic Fabric elements are the same in each case: computers, storage systems, and networks. Furthermore, each resource speaks standard Connectivity protocols for communication and security, and Resource protocols for enquiry, allocation, and management. Above this, each application uses a mix of generic and more application-specific Collective services.

Let us consider the ray tracing application in a little more detail. We assume that this is based on a high-throughput computing system [47]. In order to manage the execution of large numbers of largely independent tasks in a VO environment, this must keep track of the set of active and pending tasks, locate appropriate resources for each task, stage executables to those resources, detect and respond to various types of failure, and so forth. An implementation in the context of our Grid architecture uses both domain-specific Collective services (dynamic checkpoint, task pool management, failover) and more generic Collective services (brokering, data replication for executables and common input files), as well as standard Resource and Connectivity protocols.

Table 1: The Grid services used to construct the two example applications of Figure 1.

|  | **Multidisciplinary Simulation** | **Ray Tracing** |
|---|---|---|
| **Collective (application-specific)** | Solver coupler, distributed data archiver | Checkpointing, job management, failover, staging |
| **Collective (generic)** | Resource discovery, resource brokering, system monitoring, community authorization, certificate revocation | |
| **Resource** | Access to computation; access to data; access to information about system structure, state, performance. | |
| **Connectivity** | Communication (IP), service discovery (DNS), authentication, authorization, delegation | |
| **Fabric** | Storage systems, computers, networks, code repositories, catalogs | |

# 6   "On the Grid": The Need for Intergrid Protocols

Our Grid architecture establishes requirements for the protocols and APIs that enable sharing of resources, services, and code. It does not otherwise constrain the technologies that might be used to implement these protocols and APIs. In fact, it is quite feasible to define multiple instantiations of key Grid architecture elements. For example, we can construct both Kerberos- and PKI-based protocols at the Connectivity layer. However, Grids constructed with these different protocols are not interoperable and cannot share essential services.

The long-term success of Grid computing requires that we define and achieve widespread deployment of standard *Intergrid protocols* at the Connectivity and Resource layers—and, to a lesser extent, at the Collective layer. Much as the core Internet protocols enable different computer networks to interoperate and exchange information, these Intergrid protocols enable different organizations to interoperate and exchange or share resources. Standard APIs are also highly useful if Grid code is to be shared. The definition of these Intergrid protocols and APIs is beyond the scope of this article, although the Globus Toolkit represents an approach that has had some success to date.

We emphasize that the definition of intergrid protocols does not preclude the creation of private Grids that interoperate within well-defined boundaries. Just at intranets exploit Internet protocols within confined settings, intragrids can use Intergrid protocols within similar constrained environments. Within intragrids, we also have the potential to produce more specialized infrastructures in which key protocols are substituted. For example, let us assume that, as advocated in [33], the Intergrid protocols are PKI based. We might nevertheless substitute Kerberos for PKI within a closed environment in which Kerberos can be assumed [10]; similarly, we might use a specialized transport protocol in an embedded grid that operates entirely within a system area network. We would not say that resources within these grids were "on the Grid," however, just as we would not say that a resource that speaks something other than the Internet protocols is "on the Net."

The importance of standard protocols and services is not always clear to tool developers, who historically have often developed vertically integrated solutions, defining, for example, application-specific resource access services, information access services, and security services. This approach is not tenable for the long term, however. The costs associated with deploying and operating a resource access, information, or security service precludes sites maintaining multiple instances of such services on a large scale. The Grid community must agree on standards in these and other areas, or it will fail in its goal of broad adoption of its technologies.

# 7   Relationships with Other Technologies

The concept of controlled, dynamic sharing within VOs is so fundamental that we might assume that Grid-like technologies must surely already be widely deployed. In practice, however, while the need for these technologies is indeed widespread, in a wide variety of different areas we find only primitive and inadequate solutions to VO problems. In brief, current distributed computing approaches do not provide a general resource-sharing framework that addresses VO requirements. Grid technologies distinguish themselves by providing this generic approach to resource sharing. This situation points to numerous opportunities for the application of Grid technologies.

## 7.1   World Wide Web

The ubiquity of Web technologies (i.e., IETF and W3C standard protocols—TCP/IP, HTTP, SOAP, etc.—and languages, such as HTML and XML) makes them attractive as a platform for constructing VO systems and applications. However, while these technologies do an excellent job of supporting the browser-client-to-web-server interactions that are the foundation of today's Web, they lack features required for the richer interaction models that occur in VOs. For example, today's Web browsers typically use TLS for authentication, but do not support single sign-on or delegation.

Clear steps can be taken to integrate Grid and Web technologies. For example, the single sign-on capabilities provided in the GSI extensions to TLS would, if integrated into Web browsers, allow for single sign-on to multiple Web servers. GSI delegation capabilities would permit a browser client to delegate capabilities to a Web server so that the server could act on the client's behalf. These capabilities, in turn, make it much easier to use Web technologies to build "VO Portals" that provide thin client interfaces to sophisticated VO applications. WebOS addresses some of these issues [58].

## 7.2   Application and Storage Service Providers

Application service providers, storage service providers, and similar hosting companies typically offer to outsource specific business and engineering applications (in the case of ASPs) and storage capabilities (in the case of SSPs). A customer negotiates a service level agreement that defines access to a specific combination of hardware and software. Security tends to be handled by using VPN technology to extend the customer's intranet to encompass resources operated by the ASP or SSP on the customer's behalf. Other SSPs offer file-sharing services, in which case access is provided via HTTP, FTP, or WebDAV with user ids, passwords, and access control lists controlling access.

From a VO perspective, these are primitive technologies. VPNs and static configurations make many VO sharing modalities hard to achieve. For example, it is typically impossible for an ASP application to access data located on storage managed by a separate SSP. Similarly, dynamic reconfiguration of resources within a single ASP or SPP is challenging and, in fact, is rarely attempted. The load sharing across providers that occurs on a routine basis in the electric power industry is unheard of in the hosting industry. A basic problem is that a VPN is not a VO: it cannot extend dynamically to encompass other resources and does not provide the remote resource provider with any control of when and whether to share its resources.

The integration of Grid technologies into ASPs and SSPs can enable a much richer range of possibilities. For example, standard Grid services and protocols can be used to achieve a decoupling of the hardware and software. A customer could negotiate an SLA for particular hardware resources and then use Grid resource protocols to dynamically provision that hardware to run customer-specific applications. A customer application running on an ASP computer could directly, efficiently, and securely use SSP storage—and/or couple resources from multiple ASPs and SSPs with their own resources, when required for more complex problems. A single sign-on security infrastructure able to span multiple security domains dynamically is, realistically, required to support such scenarios. Grid resource management and accounting/payment protocols that allow for dynamic provisioning and reservation of capabilities (e.g., amount of storage, transfer bandwidth, etc.) are also critical.

## 7.3 Enterprise Computing Systems

Enterprise development technologies such as CORBA, Enterprise Java Beans, Java 2 Enterprise Edition, and DCOM are all systems designed to enable the construction of distributed applications. They provide standard resource interfaces, remote invocation mechanisms, and trading services for discovery and hence make it easy to share resources within a single organization. However, these mechanisms address none of the specific VO requirements listed above. Sharing arrangements are typically relatively static and restricted to occur within a single organization. The primary form of interaction is client-server, rather than the coordinated use of multiple resources.

These observations suggest that there should be a role for Grid technologies within enterprise computing. For example, in the case of CORBA, we could construct an object request broker (ORB) that uses GSI mechanisms to address cross-organizational security issues. We could implement a Portable Object Adaptor that speaks the Grid resource management protocol to access resources spread across a VO. We could construct Grid-enabled Naming and Trading services that use Grid information service protocols to query information sources distributed across large VOs. In each case, the use of Grid protocols provides enhanced capability (e.g., interdomain security) and enables interoperability with other (non-CORBA) clients. Similar observations can be made about Java and Jini. For example, Jini's protocols and implementation are geared toward a small collection of devices. A "Grid Jini" that employed Grid protocols and services would allow the use of Jini abstractions in a large-scale, multi-enterprise environment.

## 7.4 Internet and Peer-to-Peer Computing

Peer-to-peer computing (as implemented, for example, in the Napster, Gnutella, and Freenet [23] file sharing systems) and Internet computing (as implemented, for example by the SETI@home, Parabon, and Entropia systems) is an example of the more general ("beyond client-server") sharing modalities and computational structures that we referred to in our characterization of VOs. As such, they have much in common with Grid technologies.

In practice, we find that the technical focus of work in these domains has not overlapped significantly to date. One reason is that peer-to-peer and Internet computing developers have so far focused entirely on vertically integrated ("stovepipe") solutions, rather than seeking to define common protocols that would allow for shared infrastructure and interoperability. (This is, of course, a common characteristic of new market niches, in which participants still hope for a monopoly.) Another is that the forms of sharing targeted by various applications are quite limited, for example, file sharing with no access control, and computational sharing with a centralized server.

As these applications become more sophisticated and the need for interoperability becomes clearer we will see a strong convergence of interests between peer-to-peer, Internet, and Grid computing [28]. For example, single sign-on, delegation, and authorization technologies become important when computational and data sharing services must interoperate, and the policies that govern access to individual resources become more complex.

## 8 Other Perspectives on Grids

The perspective on Grids and VOs presented in this article is of course not the only view that can be taken. We summarize here—and critique—some alternative perspectives (given in italics).

*The Grid is a next-generation Internet.* "The Grid" is not an alternative to "the Internet": it is rather a set of additional protocols and services that build on Internet protocols and services to support the creation and use of computation- and data-enriched environments. Any resource that is "on the Grid" is also, by definition, "on the Net."

*The Grid is a source of free cycles.* Grid computing does not imply unrestricted access to resources. Grid computing is about controlled sharing. Resource owners will typically want to enforce policies that constrain access according to group membership, ability to pay, and so forth. Hence, accounting is important, and a Grid architecture must incorporate resource and collective protocols for exchanging usage and cost information, as well as for exploiting this information when deciding whether to enable sharing.

*The Grid requires a distributed operating system.* In this view, Grid software should define the operating system services to be installed on every participating system, with these services providing for the Grid what an operating system provides for a single computer: namely, transparency with respect to location, naming, security, and so forth. Put another way, this perspective views the role of Grid software as defining a virtual machine. However, we feel that this perspective is inconsistent with our primary goals of broad deployment and interoperability. We argue that the appropriate model is rather the Internet Protocol suite, which provides largely orthogonal services that address the unique concerns that arise in networked environments. The tremendous physical and administrative heterogeneities encountered in Grid environments means that the traditional transparencies are unobtainable; on the other hand, it does appear feasible to obtain agreement on standard protocols. We define the smallest possible set of protocols that a resource *must* speak to be "on the Grid"; beyond this, we seek only to provide a framework within which many behaviors can be specified. That is, we are open rather than prescriptive.

*The Grid requires new programming models.* Programming in Grid environments introduces challenges that are not encountered in sequential (or parallel) computers, such as multiple administrative domains, new failure modes, and large variations in performance. However, we argue that these are incidental, not central, issues and that the basic programming problem is not fundamentally different. As in other contexts, abstraction and encapsulation can reduce complexity and improve reliability. But, as in other contexts, it is desirable to allow a wide variety of higher-level abstractions to be constructed, rather than enforcing a particular approach. So, for example, if a developer which believes that a universal distributed shared memory model can simplify Grid application development should implement this model in terms of Grid protocols, extending or replacing those protocols only if they prove inadequate for this purpose. Similarly, a developer who believes that all Grid resources should be presented to users as objects needs simply to implement an object-oriented "API" in terms of Grid protocols.

*The Grid makes high-performance computers superfluous.* The hundreds, thousands, or even millions of processors that may be accessible within a VO represent a significant source of computational power, if they can be harnessed in a useful fashion. This does not imply, however, that traditional high-performance computers are obsolete. Many problems require tightly coupled computers, with low latencies and high communication bandwidths; Grid computing is likely to increase, rather than reduce, demand for such systems by making access easier.

# 9   Summary

We have provided in this article a concise statement of the "Grid problem," which we define as controlled and coordinated resource sharing and resource use in dynamic, scalable virtual organizations. We have also presented both requirements and a framework for a Grid architecture, identifying the principal functions required to enable sharing within VOs and defining key relationships among these different functions. Finally, we have discussed in some detail how Grid technologies relate to other important technologies.

We hope that the vocabulary and structure introduced in this document will prove useful to the emerging Grid community, by improving understanding of our problem and providing a common language for describing solutions. We also hope that our analysis will help establish connections among Grid developers and proponents of related technologies.

The discussion in this paper also raises a number of important questions. What are appropriate choices for the Intergrid protocols that will enable interoperability among Grid systems? What services should be present in a persistent fashion (rather than being duplicated by each application) to create usable Grids? And what are the key APIs and SDKs that must be delivered to users in order to accelerate development and deployment of Grid applications? We have our own opinions on these questions [33], but the answers clearly require further research.

# Acknowledgments

We are grateful to numerous colleagues for discussions on the topics covered here, in particular Randy Butler, Steve Fitzgerald, Bill Johnston, Miron Livny, Reagan Moore, Rick Stevens, and Gregor von Laszewski, and participants in the workshop on Clusters and Computational Grids for Scientific Computing (Lyon, September 2000) and the 4$^{th}$ Grid Forum meeting (Boston, October 2000), at which early versions of these ideas were presented.

This work was supported in part by the Mathematical, Information, and Computational Sciences Division subprogram of the Office of Advanced Scientific Computing Research, U.S. Department of Energy, under Contract W-31-109-Eng-38; by the Defense Advanced Research Projects Agency under contract N66001-96-C-8523; by the National Science Foundation; and by the NASA Information Power Grid program.

# Appendix: Definitions

We define here four terms that are fundamental to the discussion in this article but are frequently misunderstood and misused, namely, protocol, service, SDK, and API.

Protocol.  A *protocol* is a set of rules that end points in a telecommunication system use when exchanging information. For example:

- The Internet Protocol (IP) defines an unreliable packet transfer protocol.

- The Transmission Control Protocol (TCP) builds on IP to define a reliable data delivery protocol.

- The Transport Layer Security (TLS) Protocol [26] defines a protocol to provide privacy and data integrity between two communicating applications. It is layered on top of a reliable transport protocol such as TCP.
- The Lightweight Directory Access Protocol (LDAP) builds on TCP to define a query-response protocol for querying the state of a remote database.

An important property of protocols is that they admit to multiple implementations: two end points need only implement the same protocol to be able to communicate. Standard protocols are thus fundamental to achieving *interoperability* in a distributed computing environment.

A protocol definition also says little about the behavior of an entity that speaks the protocol. For example, the FTP protocol definition indicates the format of the messages used to negotiate a file transfer but does not make clear how the receiving entity should manage its files.

As the above examples indicate, protocols may be defined in terms of other protocols.

Service. A *service* is a network-enabled entity that provides a specific capability, for example, the ability to move files, create processes, or verify access rights. A service is defined in terms of the protocol one uses to interact with it and the behavior expected in response to various protocol message exchanges (i.e., "service = protocol + behavior."). A service definition may permit a variety of server implementations. For example:

- An FTP server speaks the File Transfer Protocol and supports remote read and write access to a collection of files. Different FTP server implementations say support different behaviors. For example, one may allow access to files on the server's disk, another may allow access to files on a mass storage system, and another may perform caching of files in memory to improve performance under certain conditions.
- An LDAP server speaks the LDAP protocol and supports response to queries. One LDAP server implementation may respond to queries using a database of information, while another may respond to queries by dynamically making SNMP calls to generate the necessary information on the fly.

A service may or may not be persistent (i.e., always available), be able to detect and/or recover from certain errors; run with privileges, and/or have a distributed implementation for enhanced scalability. If variants are possible, then *discovery* mechanisms that allow a client to determine the properties of a particular instantiation of a service are important.

SDK. The term software development kit (SDK) denotes a set of code designed to be linked with, and invoked from within, an application program to provide specified functionality. Some SDKs provide access to services via a particular protocol. For example:

- The OpenLDAP release includes an LDAP client SDK, which contains a library of functions that can be used from a C or C++ application to perform queries to an LDAP service.
- JNDI is a Java SDK, which contains functions that can be used to perform queries to an LDAP service.

There may be multiple SDKs, for example from multiple vendors, which implement a particular protocol. Further, for client-server oriented protocols, there may be separate client SDKs for use by applications that want to access a service, and server SDKs for use by service implementers that want to implement particular, customized service behaviors.

An SDK need not speak any protocol. For example, an SDK that provides numerical functions may act entirely locally and not need to speak to any services to perform its operations.

API. An Application Program Interface (API) defines a standard interface (e.g., set of subroutine calls, or objects and method invocations in the case of an object-oriented API) for invoking a specified set of functionality. For example:

- The Generic Security Service (GSS) API [45] defines standard functions for verifying identify of communicating parties, encrypting messages, and so forth.

- The Message Passing Interface API [40] defines standard interfaces, in several languages, to functions used to transfer data among processes in a parallel computing system.

An API may define multiple language bindings or use an Interface Definition Language. The language may be a conventional programming language such as C or Java, or it may be a shell interface. In the latter case, the API refers to particular a definition of command line arguments to the program, the input and output of the program, and the exit status of the program.

An API normally will specify a standard behavior but can admit to multiple implementations. In other words, there may be multiple SDKs that implement the same API.

It is important to understand the relationship between APIs and protocols. A protocol definition says nothing about the APIs that might be used to generate protocol messages. A single protocol may have many APIs; a single API may have multiple implementations that target different protocols. In brief, standard APIs enable *portability*; standard protocols enable *interoperability*. For example, both public key and Kerberos bindings have been defined for the GSS-API. Hence, a program that uses GSS-API calls for authentication operations can operate in either a public key *or* a Kerberos environment without change. On the other hand, if we want a program to operate in a public key *and* a Kerberos environment at the same time, then we need something new: a standard protocol that supports interoperability of these two environments. See Figure 5.

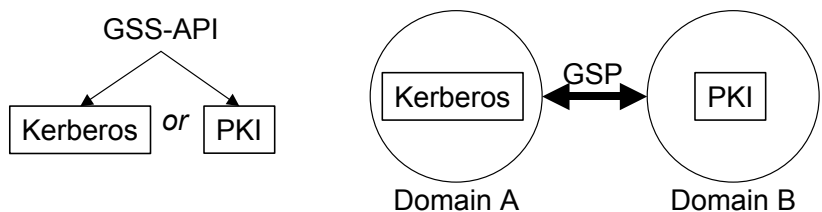

**Figure 5:** On the left, an API is used to develop applications that can target either Kerberos or PKI security mechanisms. On the right, protocols (the Grid Security Protocols) are used to enable interoperability between Kerberos and PKI domains.